\documentclass[usenatbib]{mnras}

\usepackage{graphicx}
\usepackage{dcolumn}
\usepackage{amssymb} 
\usepackage{amsmath}
\usepackage{ctable}
\usepackage{float}
\usepackage{bm}
\usepackage{float}
\usepackage{epsfig}
\usepackage{epstopdf}
\usepackage{epsf,color}
\usepackage{comment}
\usepackage{aas_macros}
\usepackage{url}
\usepackage{widetext}
\usepackage{color, colortbl}
\newcommand{\cmnt}[1]{}

\usepackage[normalem]{ulem}
\usepackage{color}
\newcommand{\repl}[2]{\sout{#1}\textcolor{red}{#2}}
\newcommand{\erscomment}[1]{\textcolor{green}{#1}}

\newcommand{\red}{\textcolor{red}}

\usepackage{tgtermes}
\newcommand{\cii}{\ion{C}{ii}}
\newcommand{\oi}{\ion{O}{i}}
\newcommand{\oiii}{\ion{O}{iii}}
\newcommand{\nii}{\ion{N}{ii}}

\title[$\cii$ evidence]{Evidence for $\cii$ diffuse line emission at redshift $z\sim2.6$}

\author[S. Yang et al.]{
Shengqi Yang,$^{1}$\thanks{E-mail:sy1823@nyu.edu}
Anthony R. Pullen,$^{1}$
Eric R. Switzer$^{2}$
\\
$^{1}$Center for Cosmology and Particle Physics, Department of physics, New York University, 726 Broadway, New York, NY, 10003, U.S.A.\\
$^{2}$NASA Goddard Space Flight Center, Greenbelt, MD 20771, USA
}

\date{Accepted XXX. Received YYY; in original form ZZZ}

\pubyear{2018}

\begin{document}
\label{firstpage}
\pagerange{\pageref{firstpage}--\pageref{lastpage}}
\maketitle

\begin{abstract}
$\cii$ is one of the brightest emission lines from star-forming galaxies and is an excellent tracer for star formation. Recent work measured the $\cii$ emission line amplitude for redshifts $2<z<3.2$ by cross-correlating Planck High Frequency Instrument emission maps with tracers of overdensity from the Baryon Oscillation Spectroscopic Sky Survey, finding $I_{\cii}=6.6^{+5.0}_{-4.8}\times 10^4$\,Jy/sr at $95\%$ confidence. In this paper, we present a refinement of this earlier work by improving the mask weighting in each of the Planck bands and the precision in the covariance matrix. We report a detection of excess emission in the 545\,GHz Planck band separate from the cosmic infrared background (CIB) present in the 353-857\,GHz Planck bands.  This excess is consistent with redshifted \cii\ emission, in which case we report $b_{\cii}I_\cii =2.0^{+1.2}_{-1.1}\times 10^5$\,Jy/sr at $95\%$ confidence, which strongly favors many \textit{collisional excitation} models of $\cii$ emission. Our detection shows strong evidence for a model with a non-zero $\cii$ parameter, though line intensity mapping observations at high spectral resolution will be needed to confirm this result.
\end{abstract}




\section{Introduction}
The star formation rate steadily increases after the first galaxies form and then dramatically declines by a factor of 20 from $z\sim 3$ to the present. One key to solve this quenching puzzle may be the interstellar medium (ISM), which provides the birthplace of stars and plays a crucial role in galaxy evolution. Studies of the molecular and fine structure lines emitted from different phases of ISM are particularly useful to unveil the ISM properties during the epoch of interest \citep{2013ARA&A..51..105C}.\par
$\cii$, the fine structure line from ionized carbon, is a strong tracer of star formation. Since the ionization energy of carbon $E_{\cii}=11.26$\,eV is less than the 13.6\,eV required to ionize hydrogen, ionized carbon is abundant under a wide variety of conditions. When the gas temperature is higher than 91\,K, $\cii$ is excited through the $^2P_{3/2}\rightarrow ^2P_{1/2}$ transition, which produces the $\cii$ emission line at $157.7\,\mathrm{\mu m}$. $\cii$ is the brightest far-infrared line, contributing $0.1-1\%$ of the total far-infrared luminosity of the nuclear region of galaxies, and has been successfully detected out to redshift 7 by the Atacama Large Microwave Submillimeter Array (ALMA) \citep{2017ApJ...836L...2B}. However, traditional galaxy redshift surveys have limitations. Firstly, surveys at high redshift tend only to resolve the brightest sources, resulting in a sample that is not representative of the average galaxy population \citep{2015ApJ...803...34B}. On the other hand, surveys with small area at low redshift may have high cosmic variance. Secondly, since high flux sensitivity drives large apertures, spectroscopic surveys for individual galaxies are expensive. Instead of resolving individual objects, we pursue an emerging approach known as intensity mapping (IM). IM is a blind and unbiased measurement. It integrates the emission along the line of sight from all sources, so it can capture faint sources across large volumes and only requires modest aperture sizes. IM was originally developed to study 21\,cm radiation from reionization but has been applied to mapping other bright lines \citep{1979MNRAS.188..791H,1990MNRAS.247..510S,1997ApJ...475..429M,1999ApJ...512..547S,2008MNRAS.383.1195W,2017arXiv170909066K,2008PhRvL.100i1303C}. 

\cite{2018MNRAS.478.1911P}, hereafter AP2018, sought to measure the intensity of $\cii$ cumulative emission through cross-correlating intensity maps with other tracers of large-scale structure (LSS). Although AP2018 reported a $\cii$ intensity brightness at redshift $z\sim2.6$, Bayesian analysis did not show a strong preference for an emission model that requires $\cii$ versus one without $\cii$. To simplify the mode-coupling matrix calculation for the angular power spectrum measurement under partial sky coverage, AP2018 used the apodized product of the Planck and galaxy surveys binary masks as a common weight for all maps. This map weighting scheme simplifies the estimator but makes it less optimal in two ways: (1) AP2018 neglects the differences of the coverage between Planck $\times$ quasar (QSO) and the Planck $\times$ CMASS luminous red galaxy (LRG) when measuring the auto-power spectra of the Planck intensity maps, which are used to construct the covariance matrix, and (2) the binary mask does not reflect the variation in noise across the Planck or LSS survey.

In this letter, we follow the measurement method proposed by AP2018, but employ more optimal map weighting. We detect an anomalous intensity consistent with $\cii$ line emission at $z\sim2.6$ with the constraint $b_{\cii}I_\cii =2.0^{+1.2}_{-1.1}\times 10^5$\,Jy/sr at $95\%$ confidence, where $b_{\cii}$ is the clustering bias of the $\cii$ emitters. Our Bayesian evidence calculation shows strong preference for $\cii$ emission in the model. We also test our measurement for systematic effects and compare it to the predictions of several promising theoretical models. However, we cannot rule out an unknown extragalactic source, which also correlates with the QSO overdensity, emitting at around 1900 GHz in the rest-frame, or a CIB model with an erroneous redshift or spectral dependence; thus we do not claim a \cii\ detection. If we interpret the excess as a \cii\ detection, the \textit{collisional excitation} models are strongly preferred.\par

\section{Data}\label{s:2}

Following AP2018, we cross-correlate Planck intensity maps with LSS tracer maps and perform a Monte Carlo Markov Chain (MCMC) analysis. The Planck maps we use are from the High-Frequency Instrument (HFI) in frequency channels 353, 545 and 857\,GHz \citep{2010A&A...520A...9L,2011A&A...536A...6P}. Two LSS surveys we use are the Sloan Digital Sky Survey (SDSS) III \citep{2011AJ....142...72E} Baryon Oscillation Spectroscopic Survey (BOSS) spectroscopic quasar sample from Data Release 12 (DR12) and the CMASS spectroscopic galaxy sample from BOSS DR12 \citep{2013AJ....145...10D}. Details about the telescope and instruments of SDSS can be found in \cite{1996AJ....111.1748F,1998AJ....116.3040G,2006AJ....131.2332G,2010AJ....139.1628D,2013AJ....146...32S}. Since the rest-frame frequency of $\cii$ is $\nu_{\cii}=1901.3$\,GHz and the redshift range of the BOSS spectroscopic quasars is $z\in[2,3.2]$, $\cii$ emission will appear in the cross-correlation at frequencies $\nu\in[450,650]$\,GHz. Similarly, the $\cii$ signal in which the Planck maps cross-correlate with CMASS galaxies at redshift $[0.43, 7]$ would be in the range $[1118,1330]$\,GHz, which is not covered by any Planck band. Therefore, $\cii$ emission only appears in one cross-correlation pair between the 545\,GHz Planck map and the BOSS QSO density field. Data from the other five cross-correlation angular power spectra are used to fit for the CIB parameters.\par
In order to test the reliability of our correlated CIB model and gain support for the $\cii$ intensity, we measure the cross-power between the 545\,GHz Planck map and a third LSS survey, which does not participate in the parameter fitting process. The LSS survey we use is the SDSS-IV \citep{2017AJ....154...28B,2018ApJS..235...42A} extended Baryon Oscillation Spectroscopic Survey (eBOSS) spectroscopic quasar sample \citep{2016AJ....151...44D}, which covers 3358\, $\mathrm{deg^2}$ of the sky and contains 80115 quasars after masking. The redshift range of eBOSS QSO is $z\in[0.8,2.0]$.\par

We multiply the Planck Galactic emission mask with the combined Planck point source mask to get the base Planck mask $W$. To assign pixels that better reflect the Planck survey depth, we modify the base Planck mask in each frequency band as
    $W_i=W(\frac{1}{\langle H_i\rangle}+\frac{1}{H_i})^{-1}\,$ ,
here $i=353,545,857$ specify the frequency, and $H_i$ is the hits counting map for the corresponding frequency channel. Instead of using the hits maps directly as weights, this harmonic mean form of weighting trades some optimality for keeping the mode-mode coupling manageable. \par

\section{Method}\label{s:3}

The method we use to measure the angular power spectra and to compute the covariance matrix follows from AP2018. We therefore refer the reader to AP2018 for details. As a brief review, we use the pseudo-$C_\ell$ estimator to measure the angular power spectra from the masked maps. Unlike in AP2018, the Planck masks we use are weighted by the hits maps, as introduced in Section \ref{s:2}. We use the 6 measured cross-power spectra between Planck intensity maps and LSS maps $C_\ell^{\{353, 545, 857\}\times\{\mathrm{QSO},\mathrm{LRG}\}}$ as data in the likelihood. We also measure the auto correlations to construct the covariance matrix. Following \cite{2002ApJ...567....2H}, we compute the mode-mode coupling matrices $M_{\ell\ell'}$ for all the possible mask pair combinations, which are important for an accurate covariance matrix calculation. This step is the most consequential change compared to AP2018.\par
  We consider the multipole range $100\leq\ell\leq1000$ and equally bin the measurement into 9 bins, each with bin width $\Delta b=100$.
We bin the mode-mode coupling matrices and the measured angular power spectra, then use the measured angular power spectra to interpolate the continuous $C_\ell$ and analytically compute the covariance matrix following \cite{2005MNRAS.358..833T}. To ensure that the covariance matrix is symmetric, we follow the treatment suggested by \cite{2005MNRAS.360.1262B}, shown in Eq (11) on AP2018.

\section{Results}\label{s:4}
Using the measured angular power spectra and the inverse covariance matrix as inputs, we perform a Monte Carlo Markov Chain (MCMC) exploration on model introduced in AP2018 section 5.1. Throughout this paper we set the minimum halo mass as $M_{\mathrm{min}}=10^{10}\,M_{\odot}$ for the MCMC fitting \citep{2013ApJ...772...77V, Serra:2014pva}. 
We vary the six parameters $\{L_0, \delta, T_{\mathrm{dust}},A_{\mathrm{tSZ}}, A_{\mathrm{exc}}, b_{\mathrm{QSO}}\}$, that float freely during the fitting process. We introduce these parameters below. \par 
The angular cross-power spectrum between Planck intensity mapping and LSS tracer map $C_\ell^{\mathrm{T-LSS}}$ is modeled as Eq (12) on AP2018. The clustering bias for BOSS QSO sample $b_{\mathrm{QSO}}$ is a free parameter floating in range $3.2-3.8$ \citep{2012MNRAS.424..933W} in the model. We use \textsc{CAMB} \citep{2000ApJ...538..473L} to compute the dark matter power spectrum in this work. For the Planck CIB source, the clustering bias $b_{\mathrm{CIB}}(k,z)$ and redshift distribution $dS/dz$ are all predicted using the halo model introduced by \cite{2012MNRAS.421.2832S}, shown in Eq (13)-(14) on AP2018. The luminosity amplitude parameter $L_0$, redshift evolution parameter $\delta$, dust temperature averaged over the redshift range $T_{\mathrm{dust}}$ are all introduced to describe the galaxy luminosity $L_{\nu(1+z)}$. The galaxy spectral energy distribution (SED) $\Theta(\nu)$, which is the frequency factor in $L_{\nu(1+z)}$, is modified as $\Theta(\nu)(1+A_{\mathrm{exc}}\delta(\nu - \nu_{\cii}))$ to account for the excess of the angular cross-power spectrum due to the $\cii$ emission. Notice that we define $C_\ell^{\cii-\mathrm{LSS}}\propto b_{\mathrm{CIB}}A_{\mathrm{exc}}=b_{\cii}(\frac{b_{\mathrm{CIB}}}{b_\cii}A_{\mathrm{exc}})=b_{\cii}A_{\cii}$ in the model, and the intensity $I$ is proportional to the amplitude parameter $A$, the intensity of $\cii$ emission $I_\cii=\frac{b_{\mathrm{CIB}}}{b_\cii}I_{\mathrm{exc}}$. Assuming the clustering bias of the CIB and $\cii$ sources are identical, $I_\cii=I_{\mathrm{exc}}$. Finally, an amplitude parameter $A_{tSZ}$ describes the contribution of correlated thermal Sunyaev-Zeldovich (tSZ) emission \citep{1972CoASP...4..173S}, at 353\,GHz.
\par
\begin{figure}
    \centering
    \includegraphics[width=0.45\textwidth]{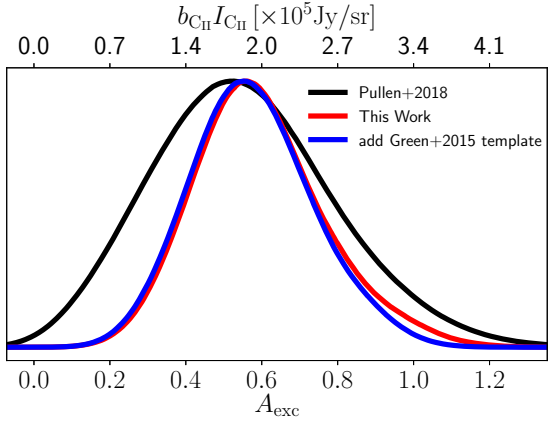}
    \caption{Posterior of the $\cii$ emission line intensity parameter $A_{\mathrm{exc}}$ from \textsc{MCMC}.}\label{fig:2}
\end{figure}

We perform the MCMC exploration of the parameter space with a modified version of \textsc{CosmoMC} \citep{2002PhRvD..66j3511L}. The reduced $\chi^2$ for our fitting is $\chi^2/N_{d.o.f.}=\chi^2/61=1.6$, while if we remove $A_\cii$ from the model, i.e. the $A_\cii$ parameter is fixed to zero, the reduced $\chi^2$ increases to 1.9. We show the posterior of parameter $A_{\mathrm{exc}}$ in Figure \ref{fig:2}. The model constrains $A_{\mathrm{exc}}=0.59^{+0.37}_{-0.33}$ at a $95\%$ confidence level, and a $\cii$ line intensity $b_{\cii}I_\cii =2.0^{+1.2}_{-1.1}\times 10^5$\,Jy/sr.  If we assume $b_{\cii}=b_{\mathrm{CIB}}$, which we set to 2.92 at $z=2.6$, we find $I_{\cii}=6.9^{+4.2}_{-3.8}\times 10^4$\,Jy/sr, consistent with the value from AP2018 but with lower uncertainty. The best-fit values for other parameters are $L_0=0.245^{+0.040}_{-0.036}$, $\delta=2.37^{+0.19}_{-0.21}$, $T_{\mathrm{dust}}=28.3^{+1.3}_{-1.4}$\,K, $A_{\mathrm{tSZ}}=0.78^{+0.44}_{-0.47}$ and $b_{\mathrm{QSO}}=3.32^{+0.32}_{-0.13}$ at 95\% c.l.\par 
There are three other parameters used in the theoretical $C_\ell^{\mathrm{T-LSS}}$ model: $\beta$ introduced as the emissivity in the SED, and $M_{\mathrm{eff}}$, $\sigma^2_{L/m} $ used in the log-normal dark matter halo mass dependence of the galaxy luminosity. We find these three parameters have descending impact on the $C_\ell^{\mathrm{T-LSS}}$. Following AP2018, we fix these three parameters as: $\log_{10}(M_{\mathrm{eff}})[M_{\odot}]=12.6$, $\sigma^2_{L/m}=0.5$ and $\beta=1.5$, but even if we let $\beta$ or $M_{\mathrm{eff}}$ float in the MCMC process, the fitting result for $A_{\mathrm{exc}}$ is almost unchanged. \par 
Our fitting results differ from AP2018 mainly because AP2018 uses the product of the Planck mask and the galaxy survey mask as the final mask to compute $\hat{C}_b^{\mathrm{TL}}$. When computing $Cov[\hat{C}_b^{\mathrm{T}\times \mathrm{QSO}},\hat{C}_{b'}^{\mathrm{T}\times \mathrm{LRG}}]$, which requires the auto-correlation of Planck maps, AP2018 assumes that CMASS galaxy mask and BOSS QSOs mask are similar and use the product of the Planck mask and BOSS QSOs mask to measure $C_\ell^{\mathrm{TT}}$ for simplicity. Since the mode-mode coupling matrices are computationally expensive, this approach is faster but reduces optimality of the estimator. In this work, we distinguish the weight for each map and compute all the mode-mode coupling matrices that are needed to keep the covariance matrix as accurate as possible. As a result, the $A_{\mathrm{exc}}$ measurement does not shift much, but the standard deviation in this work decreases by a factor of 1.4.\par


\section{Tests of detection}\label{s:5}
\subsection{Bayesian Analysis}
To test if this non-zero excess emission is a true detection, we use Bayesian evidence with the Laplace approximation to determine if introducing the parameter $A_{\mathrm{exc}}$ into the model is preferred:
\begin{equation}\label{eq:1}
    \langle B\rangle=\dfrac{1}{\sqrt{2\pi}}\dfrac{\sqrt{det F_{\mathrm{exc}}}}{\sqrt{det F}}exp\left[-\dfrac{1}{2}\delta\theta_\alpha F^{\alpha\beta}_{\mathrm{exc}}\delta\theta_\beta\right]\Delta\theta_{A_{\mathrm{exc}}}\, .
\end{equation}
Here $F_{\mathrm{exc}}$ and $F$ are Fisher matrices for model with and without $A_{\mathrm{exc}}$, respectively. $\delta\theta$ is the parameter fitting difference between the two models in comparison, and $\Delta\theta_{A_{\mathrm{exc}}}$ is the prior of $A_{\mathrm{exc}}$. We find the Bayesian evidence $\langle B\rangle=0.066$, which shows a strong preference to the model with a free $A_{\mathrm{exc}}$ parameter.\par
Following \citet{2018arXiv181206223S}, hereafter S2019, which consider the cross-power with the \citet{2015ApJ...810...25G} Milky Way template to provide additional leverage to suppress foregrounds, we add a term to the model which can accommodate correlations between the \citet{2015ApJ...810...25G} model and LSS due to systematics. This is parameterized as amplitude $\alpha$ times the CIB clustering anisotropy template. CosmoMC fits (Figure \ref{fig:2}) $A_{\mathrm{exc}}=0.57^{+0.33}_{-0.31}$ under S2019 model, corresponding to a mean intensity of $b_{\cii}I_{\cii}=2.0^{+1.0}_{-1.1}\times10^5 $\,Jy/sr ($95\%$ c.l.). The Bayesian evidence between the S2019 models with/without $A_{\mathrm{exc}}$ parameter is $\langle B\rangle=0.030$, which shows an even stronger preference to $\cii$ emission line detection. We then compare the AP2018 model with the S2018 model, both with a free $A_{\mathrm{exc}}$ parameter, but the S2019 model contains an extra parameter $\alpha$. Replacing $F_{\mathrm{exc}}$ and $\Delta\theta_{A_{\mathrm{exc}}}$ in Eq (\ref{eq:1}) by the Fisher matrix of S2019 model and the prior of parameter $\alpha$, we get $\langle B\rangle=11.29$, showing a strong preference to the AP2018 model. This indicates that with current data, projecting out foregrounds using the \citet{2015ApJ...810...25G} template is not enough to compensate the over-fitting effect caused by introducing an extra parameter. We therefore report the fitting result $b_{\cii}I_\cii =2.0^{+1.2}_{-1.1}\times 10^5$\,Jy/sr and $I_{\cii}=6.9^{+4.2}_{-3.8}\times 10^4$\,Jy/sr as our final conclusion in this work.\par

\subsection{Statistical Tests}
Spectral line contamination can bias the measurement. Besides $\cii$, other fine-structure emission lines such as $\oi$\,(145\,$\mathrm{\mu}$m), $\oiii$(88\,$\mathrm{\mu}$m) and $\nii$(205\,$\mathrm{\mu}$m) will contribute to the six sets of angular power spectra.
To test how much the contamination from other lines influences the data, we include lines from Table 1 of \cite{2010JCAP...11..016V} and theoretically compute the fractional deviation of angular power spectra compared to the original $C_\ell^{\mathrm{TL}}$s under the best-fit parameters. The results are summarized in Table \ref{t:1}. $C_\ell^{\cii-\mathrm{QSO}}$ only increases by $2.5\%$, and all $C_\ell^{\mathrm{T-LSS}}$ involved in the theoretical model increase less than $3\%$. Since the measurement error of $\hat{C}_\ell$ are larger than $13\%$, deviation to $A_{\mathrm{exc}}$ caused by interloper lines are not significant in this work. For interlopers to matter at the 1$\sigma$ level, the ratio of $\cii$ to interloper lines based on \citet{2010JCAP...11..016V} would need to be overestimated by a factor of five.\par
\begin{table}
    \centering
    \begin{tabular}{c c c}
    \hline\hline
         $C_\ell$& Interloper &$\Delta C_\ell/C_\ell[\%]$  \\\hline
         353-QSO&$^{12}$CO(10-9),$^{12}$CO(11-10), &0.69\\
         &$^{12}$CO(12-11)&\\
         545-QSO&$\oi$&0.34\\
         857-QSO&$\oiii$&1.3\\
         353-CMASS&$^{12}$CO(5-4),$^{13}$CO(5-4),HCN(6-5)&2.5\\
         545-CMASS&$^{12}$CO(7-6),$^{12}$CO(8-7),CI,&1.5\\
         &$^{13}$CO(7-6),$^{13}$CO(8-7)&\\
         857-CMASS&$^{12}$CO(11-10),$^{12}$CO(12-11),$\nii$&0.55\\
         $\cii$-QSO&all interlopers&2.5\\\hline
    \end{tabular}
    \caption{Fractional contribution by interloping spectral lines in Planck-LSS and $\cii-\mathrm{QSO}$ angular power spectra. The deviation is small compared to error in this work.}
    \label{t:1}
\end{table}

The continuum foreground contamination from the Milky Way, which is not statistically isotropic, is another concern for intensity mapping measurements. The LSS tracer maps we use themselves would not have MW emission, but the selection function may be impacted by bright galactic emission. So the overdensity implied by LSS tracer maps could be influenced by the foreground emission from Milky Way Galaxy, and we would expect the data from the measurement to vary under different sky area if that is the case. To test if the measured angular power spectra converge under decreasing survey area, we replace the $40\%$ Planck Galactic emission mask with $20\%$ mask and redo the angular power spectra measurement. The difference between the original measurement and the measurement under the smaller area is well within the uncertainty. We also perform a jackknife test to check if the measured angular power spectra agree under different sky areas. We divide the product of Planck mask and QSOs (LRGs) mask into 47 (42) regions, and apply this boundary to Planck mask and QSOs (LRGs) mask when doing the cross-correlation. All the angular cross-correlations, measured when one of the jackknife regions is excluded, agree with $\Hat{C}_\ell$ data we use within the error. We further simulate 100 random Gaussian fields and cross correlate them with the three Planck intensity maps. The angular cross-correlations are all consistent with zero mean and errors that agree with the Gaussian form for the errors used here, indicating that the non-Gaussian component in the continuum foreground does not bias our cross-power or the Gaussian estimate of bandpower errors.

\subsection{Tests of the Correlated CIB Model}
A systematic error in the CIB model
could be attributed incorrectly to $\cii$ emission and still produce an overall reasonable $\chi^2$ value. To test if the best-fit CIB model is accurate, we use the best-fit CIB parameters $\{L_0, \delta, T_{\mathrm{dust}}\}$ to predict the angular cross-power spectra between Planck CIB maps in 353, 545, 857\,GHz frequency channel and the eBOSS QSO map. We then compare the theoretical prediction with measurement through a $\chi^2$ test. We use the clustering bias fitted by \cite{2017JCAP...07..017L} for eBOSS QSO . The theoretically estimated angular cross-power spectra agree with the measurements. The reduced $\chi^2=1.22$, $1.54$, $1.62$ for $C_\ell^{353 \times \mathrm{eBOSSQSO}}$, $C_\ell^{545 \times \mathrm{eBOSSQSO}}$ and $C_\ell^{857 \times \mathrm{eBOSSQSO}}$ respectively.
\par 
We further perform a MCMC on the best-fit CIB parameters. We introduce another free parameter $A_{\mathrm{null}}$ and assume $[C_\ell^{545 \times \mathrm{eBOSS QSO}}]_{\mathrm{data}}=(1+A_{\mathrm{null}})\times[C_\ell^{545\times \mathrm{eBOSS QSO}}]_{\mathrm{camb}}$ in the theoretical model, where $[C_\ell]_{\mathrm{data}}$ is the measured angular power spectrum, while $[C_\ell]_{\mathrm{camb}}$ is theoretically computed through Eq (12) of AP2018 based on the best-fit CIB parameters. We set a constant prior of $A_{\mathrm{null}}$ from $-1$ to $1$ and fit for the 7 free parameters through \textsc{CosmoMC}. We get $A_{\mathrm{null}}=0.04^{+0.18}_{-0.18}$ at $95\%$ confidence level while other CIB/$\cii$ parameters do not change significantly. Hence, these additional cross-correlation channels are fully consistent with the correlated CIB model, and do not show any excess analogous to $545$\,GHz cross quasars. Since $C_\ell^{\mathrm{CIB}\times \mathrm{eBOSSQSO}}$ is not involved in the parameter fitting process, this result brings extra support to the reliability to the CIB model. \par

\subsection{Constraints on Theoretical Models}

\begin{figure}
    \centering
    \includegraphics[width=0.45\textwidth]{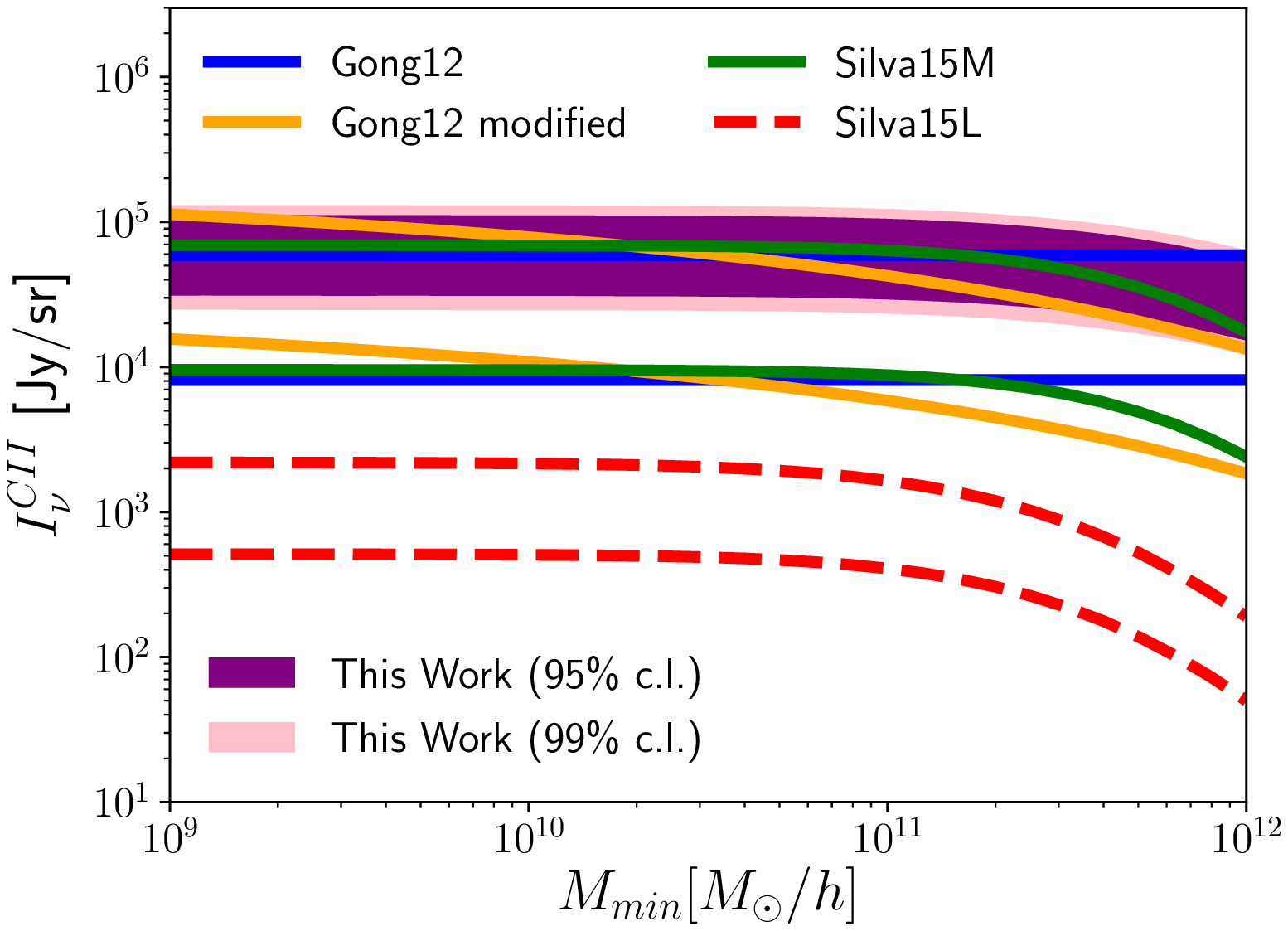}
    \caption{$\cii$ intensity measurement in this work at $95\%$ and $99\%$ c.l. and the theoretical predictions from \protect\cite{2012ApJ...745...49G, 2015ApJ...806..209S, 2018MNRAS.478.1911P} .}\label{fig:7}
\end{figure}

If we interpret the excess correlated emission of Planck $545$\,GHz cross quasars as $\cii$, we can make additional statements about the excitation of the gas. Figure \ref{fig:7} shows the $\cii$ constraint consistent with our excess measurement (assuming $b_{\mathrm{CIB}}=b_{\cii}$) together with the theoretical predictions from models \cite{2012ApJ...745...49G} (Gong12), \cite{2015ApJ...806..209S} (Silva15) and the modified Gong12 model introduced by AP2018. We include the impact of $M_{\mathrm{min}}$ on $b_{\mathrm{CIB}}(M)$ and $n(M)$, which is the number density of halos with mass above $M$. As $M$ increases, $b_{\mathrm{CIB}}(M)$ rises and $n(M)$ falls. Since $C_\ell^{\cii-\mathrm{QSO}}\sim I_\cii b_{\mathrm{CIB}}$ and $I_\cii\propto n(M)$, increasement of $M_{\mathrm{min}}$ results in a decline of $I_\cii$. 'Gong12', 'Gong12 modified' and 'Silva15M' are \textit{collisional excitation} models. The upper bound of Figure \ref{fig:7} corresponds to an infinite spin temperature for the $\cii$ transition while the lower bound corresponds to kinetic temperature of the electron $T_k^e=100$\,K and the number density of electron $n_e=1\,\mathrm{cm}^{-3}$. 'Silva15L' models $\cii$ intensity based on measured luminosity function, the upper/lower bound is model 'm1'/'m4' in \cite{2015ApJ...806..209S} Table 1. We find our measurement favors \textit{collisional excitation} models. 'Silva15L' is lower than the lower limit of our measurement at $99\%$ c.l., thus the \textit{scaling relations} models such as 'Silva15L' are strongly disfavored by our measurement. However, our 'Silva15L' model is extrapolated from redshift $z=6$, so we do not completely rule it out.\par
\begin{figure}
    \centering
    \includegraphics[width=0.45\textwidth]{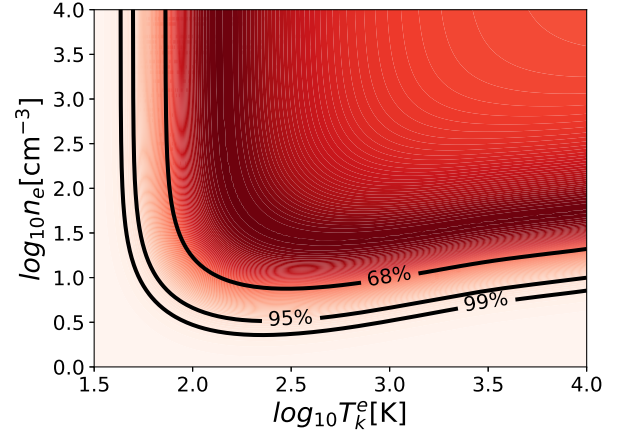}
    \caption{Contours of posterior $p[I_{\cii}(T_k^e,n_e)]$ at $68\%$, $95\%$ and $99\%$ confidence levels. The intensity of $\cii$ line $I_{\cii}$ at redshift $z=2.6$ is theoretically computed from 'Gong12 modified' model. Darker region is in the higher confidence level.} \label{fig:8}
\end{figure}

In addition, assuming an excess modeled as a $\cii$ intensity can constrain the kinetic temperature of the electron $T_k^e$ and number density of electron $n_e$ in the theoretical model 'Gong12 modified'. The contour of the $I_{\cii}$ posterior is shown in Figure \ref{fig:8}. The upper bound of 'Gong12 modified' shown in Figure \ref{fig:7} corresponds to $T_k^e\rightarrow\infty$ and $n_e\rightarrow\infty$ in the parameter space, which is within $95\%$ c.l. of our measurement. Figure \ref{fig:8} also shows the asymptotic behavior of $I_{\cii}(T_k^e,n_e)$ at very high kinetic temperature or electron number density.
Another observable is needed to break the degeneracy between $T_k^e$ and $n_e$. One possibility would be another fine structure line which also depends on ionization of the ISM. Another is a measurement of the one-point PDF of $\cii$ intensities \citep{2017MNRAS.467.2996B,2019ApJ...871...75I} which samples the full luminosity function instead of just the first moment like the intensity.

\section{Conclusions}\label{s:6}
This work is the most significant detection of an excess consistent with the $\cii$ emission line from intensity maps. We use the model proposed by AP2018: using the angular cross-power spectra between Planck intensity maps in frequency $353, 545, 857$\,GHz and both BOSS quasars and CMASS galaxies to constrain a source of emission correlated with LSS consistent with $\cii$ emission at redshift $z\sim2.6$ with $b_{\cii}I_\cii =2.0^{+1.2}_{-1.1}\times 10^5$\,Jy/sr at $95\%$ confidence level, which is strongly preferred as Bayesian evidence. We find contamination from foreground anisotropy and interloper lines are not significant compared to measurement error for the data we use. Our best-fit CIB model can also successfully predict the angular cross-power spectrum between Planck intensity maps and eBOSS quasars, supporting the reliability of our fitting results. Among the $\cii$ models considered in this work, the $\cii$ constraint favors many \textit{collisional excitation} models. We study the posterior of $\cii$ line intensity as a function of electron kinetic temperature $T_k^e$ and number density $n_e$ under the theoretical model introduced by \cite{2012ApJ...745...49G} and AP2018. Based on the $\cii$ intensity constraint consistent with this work, we can constrain the values of $T_k^e$ and $n_e$. $T_k^e$ and $n_e$ are degenerate in the modified \cite{2012ApJ...745...49G} model. Measurements of other observables such as the intensity or luminosity functions of other fine structure lines may be able to break the $T_k^e$-$n_e$ degeneracy. \par
In this paper we refrain from claiming that the excess we measure is a confident detection of $\cii$, considering we cannot rule out that our excess is due to redshift evolution of the CIB parameters, in particular the spectral index $\beta$. The resolution to this question will require upcoming line intensity mapping surveys which will be able to use high spectral resolution to discriminate between \cii\ line emission and continuum CIB emission.

\section*{Acknowledgements}

We would thank Shenglong Wang for the IT support. We also thank Michael Blanton, Jeremy Tinker, Patrick Breysse and Dou Liu for useful discussions.\par
This work is based on observations obtained with \emph{Planck} (\url{http://www.esa.int/Planck}), an ESA science mission with instruments and contributions directly funded by ESA Member States, NASA, and Canada.

Funding for SDSS-III and IV have been provided by the Alfred P. Sloan Foundation, the Participating Institutions, the National Science Foundation, and the U.S. Department of Energy Office of Science. SDSS-IV acknowledges
support and resources from the Center for High-Performance Computing at
the University of Utah. The SDSS web site is \url{www.sdss.org}.

SDSS-III and IV are managed by the Astrophysical Research Consortium for the Participating Institutions of the SDSS Collaboration including the University of Arizona, the Brazilian Participation Group, Brookhaven National Laboratory, the Carnegie Institution for Science, Carnegie Mellon University, the Chilean Participation Group, University of Florida, the French Participation Group, the German Participation Group, Harvard University, Instituto de Astrof\'isica de Canarias, the Michigan State/Notre Dame/JINA Participation Group, Johns Hopkins University, Kavli Institute for the Physics and Mathematics of the Universe (IPMU) / 
University of Tokyo, the Korean Participation Group, Lawrence Berkeley National Laboratory, Leibniz Institut f\"ur Astrophysik Potsdam (AIP),  
Max-Planck-Institut f\"ur Astronomie (MPIA Heidelberg), 
Max-Planck-Institut f\"ur Astrophysik (MPA Garching), 
Max-Planck-Institut f\"ur Extraterrestrische Physik (MPE), National Astronomical Observatories of China, New Mexico State University, New York University, University of Notre Dame, 
Observat\'ario Nacional / MCTI, Ohio State University, Pennsylvania State University, Shanghai Astronomical Observatory, United Kingdom Participation Group,
Universidad Nacional Aut\'onoma de M\'exico, University of Arizona, 
University of Colorado Boulder, University of Oxford, University of Portsmouth, Princeton University, the Spanish Participation Group, University of Utah, Vanderbilt University, University of Virginia, University of Washington, University of Wisconsin, 
Vanderbilt University and Yale University.

\bibliographystyle{mnras}
\bibliography{CII}




\label{lastpage}
\end{document}